# Numerical Evidence of Small Coherent Subsystems at Low Temperatures in *Light Harvesting Complex II*

*Jiahao Chen*

*Nancy Makri\**

School of Chemical Sciences, University of Illinois, Urbana, Illinois 61801-3760

nancy@makri.scs.uiuc.edu

**February 4, 2003**

\*To whom correspondence should be addressed. Address: A442 Chemical and Life Sciences Laboratory, 601 South Goodwin Avenue, Urbana, IL 61801–3760. Tel: (217) 333-6589. Fax: (217) 244-3186.

The extent of exciton coherence in protein–pigment complexes has significant implications for the initial light harvesting step in photosynthetic organisms. In this work we model the main antenna protein of *photosystem II*, namely *light harvesting complex II* (LHC–II), with a single–exciton Hamiltonian with sites coupled via dipole–dipole interaction, with linear coupling to a dissipative phonon bath. With appropriate parameters, Monte Carlo path integral (MCPI) results of the exciton coherence length from 1 K to 500 K show that at thermodynamic equilibrium, an exciton in LHC–II is localized mostly on single chlorophyll pigment sites, with persistent short–range coherence over the A2–B2 pair, A3–B3 pair and B1–B5–B6 triplet. Quasi–adiabatic path integral (QUAPI) calculations of the subsystems mentioned above show a smooth, incoherent relaxation towards thermodynamic equilibrium. The results obtained imply that with the exception of small coherent subsystems at cryogenic temperatures, excitons in LHC–II are more localized than in the analogous *light harvesting complex* II (LH–II) of the purple bacterium *Rs. molischianum*, which may be expected from the lower symmetry of the former.

## I. Introduction

Since the pioneering work of Réne Dutrochet in 1837 and Richard M. Willstätter in 1905, chlorophyll has been recognized as the main component of plant matter essential to photosynthesis.[1] However, the existence of chlorophyll *in vivo* as protein–pigment complexes[2,3] was only discovered in 1956 with the discovery of *photosystem I*,[4] followed by *photosystem* II in 1969.[5] The majority of chlorophyll (Chl) found in green plants is now known to be bound in the protein *light–harvesting complex* II (LHC–II), the principal antenna complex for *photosystem II*. Thus studying the highly efficient energy trapping process of light–harvesting proteins is not only relevant to uncovering the origins of the world's ecosystems,[7] but is also of prime importance for designing artificial solar energy collectors.[8–11]

The characterization of LHC–II in 1994 produced only a rough model,[12] unable to distinguish neither between chlorophylls *a* and *b* (Chl *a* and *b* respectively), nor between the *x* and *y* axes of the porphyrin macrocycle—testimony to the difficulty of isolating membrane proteins.[13–15] Biochemical plausibilities led to the assignment of Chl *a* sites A1–A7 and Chl *b* sites B1–B6 (the existence of B4 is still under consideration); the experimental resolution was insufficient to directly observe the chlorophylls beyond the position and planar orientation of the rings. Numerical simulations[16,17] and fits to experimental spectra,[18,19] in conjunction with site mutagenesis experiments,[20] reduced the uncertainty in the site identities and orientations to the sites A3, A5, B3, and B6.[20–22] LHC–II shows significant sequence homology with other light harvesting complexes such as the minor peripheral antenna complexes CP29, CP26 and CP24, belonging to photosystem II; LHC–I from photosystem I; algal Chl *a*/*c* complexes; and early–light induced proteins.[23–25] Such sequence homology is considered genetic evidence suggesting evolutionary relationships between these complexes.

Although ultrafast spectroscopic techniques allow the measurement of excitonic effects,[26,27] the exact loci of harvested excitons in protein–pigment complexes have yet to be determined. Of particular interest is the comparison of LHC–II with the analogous[28,29] LH–II complex found in purple bacteria. The B800/B850 ring for the LH–II system exhibits high symmetry of its monomers, such as $C_8$ of the B800 ring in LH–II from *Rs. molischianum*. LHC–II has a higher pigment density than in LH–II, and the presence of strong pigment–pigment interactions suggest that excitonic coupling may be a significant feature in the ultrafast spectral signature of LHC–II.[30,31] Precise knowledge of the dynamics of exciton formation and evolution has significant repercussions in terms of biologically hazardous energy loss processes such as singlet oxygen formation, and hence bears on organism's mechanisms for their containment via various quenching processes.[32] However, experimental evidence from Stark spectroscopy[33] seems to indicate that excitonic effects actually seem to be lower in LHC–II.[34]

It is now recognized that the proteinaceous backbone of protein–pigment complexes does not simply specify the relative orientations of the pigment molecules *in vivo*, it also acts as a dissipative thermal bath, whereby excess excitation energy may be quenched through exciton–phonon coupling. The anisotropic local environment also shifts the pigment sites from their equilibrium energy levels, lifting their degeneracies and resulting in inhomogeneous broadening of their spectral signals, facilitating broadband absorption. Through conformational changes, the protein backbone may also perform a regulatory function during light harvesting.[35]

## II. Theory and Parameterization

In this work we consider single excitons in a single LHC–II monomer (Figure 1), disregarding any possible effects of trimeric aggregation[36,37] as well as multiple exciton effects.[38] For comparison with previous work on LH–II,[39] a simple one–exciton Hamiltonian is used to describe the protein–pigment complex (PPC):

$$\hat{H}_s = \sum_{j=1}^{n} e_j |u_j\rangle\langle u_j| + \sum_{i=1}^{n}\sum_{j\neq i}^{n} V_{ij} |u_i\rangle\langle u_j| \qquad (2.1)$$

Here, $n$ is the number of sites $|u_j\rangle$ in the PPC, and $e_j$'s are the individual site energies of the chlorophyll sites, corresponding to the individual $Q_y$ transition energies. In this model, the PPC can be taken to be a closed ring with sites corresponding to possible sites of an exciton. The path of such an exciton through the PPC can be considered a one–dimensional ring described by a closed path of N steps represented by the system state vector $\vec{s} = \{s_1, \cdots, s_N\}$.[40]

Three different sets of site energies $e_j$ were used. The first corresponds to isoenergetic chlorophyll sites of energy 14 792 cm$^{-1}$,[41] which is unphysical but would show the maximum possible coherence independent of the choice of site energies. The second differentiates Chl $a$ (14 714 cm$^{-1}$) and Chl $b$ sites (15 365 cm$^{-1}$) (identity differentiation),[42] which is technically a more correct description but the resulting eigenenergies produces an energy level diagram that is in apparent conflict with the energies obtained from deconvolution studies of experimental spectra.[20,43–45] The third set corresponds to sites completely differentiated in energy, meant as a crude model for static disorder in LHC–II (A1, 14 970 cm$^{-1}$; A2, 14 881 cm$^{-1}$; A3, 14 837 cm$^{-1}$; A4, 14 837 cm$^{-1}$; A5, 14 993 cm$^{-1}$; A7, 14 881 cm$^{-1}$; B1,

15 038 cm$^{-1}$; A6, 14 859 cm$^{-1}$; B2, 14 993 cm$^{-1}$; B3, 15 504 cm$^{-1}$; B5, 15 444 cm$^{-1}$; B6, 15 396 cm$^{-1}$).[22] A more thorough study of the static disorder effect would require the random sampling of site energies from a statistical distribution.[46] Also, the observed temperature dependence of the spectral eigenenergies[20] spells caution when extrapolating data from cryogenic experiments to explain the light harvesting process at physiological temperatures.

In the absence of a more detailed physical description of the system, the off–diagonal exciton coupling elements $V_{ij}$ can be found using the simplest available model of exciton coupling,[47] the point dipole approximation:[48]

$$V_{ij}/cm^{-1} = \frac{5.04\, \mathbf{k}_{ij}}{r_{ij}^3} \vec{\mu}_i \cdot \vec{\mu}_j \qquad (2.2)$$

The equation above gives the coupling strength in wavenumbers. The symbol $\mathbf{k}_{ij}$, the orientation factor of the transition dipole moments, is given by:

$$\mathbf{k}_{ij} = \hat{\vec{\mu}}_i \cdot \hat{\vec{\mu}}_j - 3\left(\hat{\vec{\mu}}_i \cdot \hat{\vec{r}}_{ij}\right)\left(\hat{\vec{\mu}}_j \cdot \hat{\vec{r}}_{ij}\right) \qquad (2.3)$$

In LHC–II, there are twelve chlorophyll sites ($n = 12$) and the coupling elements have been calculated with suitable constants based on one particular set of orientations and identities.[34]

Experimental data seem to emphasize the importance of protein coupling to excitonic effects. To account for dissipation by the protein environment through the phenomenon of exciton–phonon coupling, a harmonic bath of fictitious particles is introduced into the total Hamiltonian:

$$\hat{H} = \hat{H}_s + \sum_j \frac{\hat{p}_j^2}{2m_j} + \frac{m_j w_j^2}{2}\left(x_j - \frac{c_j}{m_j w_j^2}\sum_{k=1}^n \mathbf{s}_k |u_k\rangle\langle u_k|\right)^2 \qquad (2.4)$$

The last term in the harmonic bath is a counterterm that renormalizes the potential so that features of interest in the system do not become dependent on the coupling strength.[49]

Instead of solving for all the bath coordinates $\vec{x} = \{x_i, \cdots, x_N\}$, it is only necessary to realize that only the collective properties of the entire bath are significant to the dynamics of the excitonic system.[50] The response of the bath is entirely contained in the spectral density expression:

$$J(\omega) = \frac{\pi}{2} \sum_j \frac{c_j^2}{m_j \omega_j} \delta(\omega - \omega_j) \qquad (2.5)$$

In the absence of a more detailed experimental description of the dissipative protein environment, a suitable model to employ is the standard Ohmic bath with the dimensionless Kondo parameter $\xi$ that describes the strength of the system–bath coupling:

$$J(\omega) = \hbar \xi \omega e^{-\frac{\omega}{\omega_c}} \qquad (2.6)$$

The critical frequency $\omega_c$ was chosen to be 120 cm$^{-1}$ as the experimental bandwidth reported by nonphotochemical hole–burning spectroscopy at 4K.[51] However, the use of this parameter may be complicated by the distinct non–Gaussian shape of the experimental spectra[52–54] due to static and dynamic disorder of the environment.[55]

The system–bath coupling strength, quantified by the dimensionless Kondo parameter $\xi$, can be estimated experimentally from the optical reorganization energy $E_r$,[33,56] the product of the Huang–Rhys factor $S$ and the energy spacing of the phonon bath $\Delta\omega$:

$$\xi = \frac{S \Delta\omega}{4\pi\hbar\omega_c} = \frac{E_r}{2\hbar\omega_c} \qquad (2.7)$$

The effective reorganization energy of the entire LHC–II complex due to exciton formation was estimated to be 70 cm$^{-1}$, as determined from the temperature dependence of the experimental bandwidth $\Gamma$:[57]

$$\Gamma^2 = \frac{E_r \omega}{2\pi} \coth \frac{\beta \hbar \omega}{2} + \Gamma_0^2 \qquad (2.8)$$

This gives an estimated Kondo parameter of $x = 0.3$, which is considered moderate dissipation.

With these parameters specified, we now turn to the methodology of simulation. For equilibrium statistical mechanics the imaginary time formalism is employed. Substitution of $t = -i\hbar b$ into the path integral expression for the real–time propagator[58] yields a real–valued system partition function of:[59]

$$Z[\vec{s}] = Tr\{e^{-b\hat{H}}\} = \sum_{k_N=1}^{N} \langle u_{k_N} | \hat{U} | u_{k_N} \rangle$$

$$= \sum_{k_1=1}^{N} \sum_{k_2=1}^{N} \cdots \sum_{k_N=1}^{N} \langle u_{k_N} | e^{-\Delta b \hat{H}_s} | u_{k_{N-1}} \rangle \cdots \langle u_{k_2} | e^{-\Delta b \hat{H}_s} | u_{k_1} \rangle \langle u_{k_1} | e^{-\Delta b \hat{H}_s} | u_{k_N} \rangle I[\vec{s}] \quad (2.9)$$

Here, the environmental effects giving rise to nuclear displacements are formally and completely separated from the Hamiltonian matrix elements in the site basis and are completely contained in the influence functional, which is given in the discretized form as:

$$I[\vec{s}] = \exp\left(-\sum_{k=1}^{N}\sum_{k'=1}^{N} h_{kk'} s_k s_{k'}\right) \quad (2.10)$$

The coefficients of the influence functional can be written as the following real integrals:[60]

$$\begin{aligned}
h_{kk'} &= -\frac{1}{p}\int_{-\infty}^{+\infty} \frac{J(w)}{w^2} \frac{e^{-\Delta b\hbar w(k-k')}}{1+e^{-b\hbar w}} (\cosh \Delta b\hbar w - 1) dw &, \quad 0 < k' < k < N \\
h_{kk} &= \frac{1}{2p}\int_{-\infty}^{+\infty} \frac{J(w)}{w^2} \frac{1-e^{-\Delta b\hbar w}}{1+e^{-b\hbar w}} dw &, \quad 0 < k < N \\
h_{N0} &= -\frac{1}{p}\int_{-\infty}^{+\infty} \frac{J(w)}{w^2} \frac{e^{-\hbar w\left(b-\frac{\Delta b}{2}\right)}}{1+e^{-b\hbar w}} \left(\cosh \frac{\Delta b\hbar w}{2} - 1\right) dw & \\
h_{00} = h_{NN} &= \frac{1}{2p}\int_{-\infty}^{+\infty} \frac{J(w)}{w^2} \frac{e^{-\frac{\Delta b\hbar w}{2}}}{1+e^{-b\hbar w}} dw & \\
h_{k0} &= -\frac{1}{p}\int_{-\infty}^{+\infty} \frac{J(w)}{w^2} \frac{1+e^{-\frac{\Delta b\hbar w}{2}}}{1+e^{-b\hbar w}} \cosh \frac{\Delta b\hbar w}{2} e^{-k\Delta b\hbar w} dw &, \quad 0 < k < N \\
h_{Nk} &= -\frac{1}{p}\int_{-\infty}^{+\infty} \frac{J(w)}{w^2} \frac{1+e^{-\frac{\Delta b\hbar w}{2}}}{1+e^{-b\hbar w}} \cosh \frac{\Delta b\hbar w}{2} e^{(b-k\Delta b)\hbar w} dw &, \quad 0 < k < N
\end{aligned} \quad (2.11)$$

The measure of delocalization employed in this study is the coherence length, $l$. As given in previous work by Ray and Makri,[39] $l$ is the weighted sum of lengths over paths of various degrees of delocalization. In the case of a ring aggregate the length can be given as the normalized standard deviation of the auxiliary path variables, which can then be evaluated using the Metropolis sampling method:[61]

$$l[\vec{s}] = 2Z[\vec{s}]\sqrt{\sum_{k=1}^{N}(s_k - \langle s \rangle)^2} \qquad (2.12)$$

Subsystems of the LHC–II Hamiltonian may be further investigated using the quasiadiabatic path integral (QUAPI) method.[60,] To calculate the non–Markovian effects that are characteristic of dissipative dynamics, the influence functional is explicitly split into different tensors (multidimensional arrays) that account for memory effects that span all relevant timeframes, up to a time $\hat{k}\Delta t$. The system path vectors are also discretized, with forward and backward paths discretized separately.

We define the notation $s_k^{\pm}$ as the state vector comprising the forward (–) and backward (+) paths at a propagation time $k\Delta t$. While the reduced (system–only) density matrix $r_s$ is the quantity of interest in this iterative algorithm, the propagation is most conveniently expressed in terms of the so–called augmented reduced density tensor, defined with the initial condition:

$$\mathbf{A}^{(\hat{k})}(s_0^{\pm},\cdots,s_{\hat{k}-1}^{\pm};t=0) = \langle s_0^+ | r_s(t=0) | s_0^- \rangle \qquad (2.13)$$

Truncating non–Markovian effects beyond a maximum temporal displacement factor $\hat{k}$, the propagation of $\mathbf{A}$ through a time increment $\hat{k}\Delta t$ may be described according to the equation:

$$\begin{aligned}\mathbf{A}^{(\hat{k})}(s_{k+\hat{k}}^{\pm},\cdots,s_{k+2\hat{k}-1}^{\pm};(k+\hat{k})\Delta t) &= \int\cdots\int \mathbf{T}^{(2\hat{k})}(s_k^{\pm},\cdots,s_{k+2\hat{k}-1}^{\pm}) \\ &\times \mathbf{A}^{(\hat{k})}(s_{k+\hat{k}}^{\pm},\cdots,s_{k+2\hat{k}-1}^{\pm};(k+\hat{k})\Delta t)\mathrm{d}s_k^{\pm}\cdots\mathrm{d}s_{k+\hat{k}-1}^{\pm}\end{aligned} \qquad (2.14)$$

The propagation tensor $\mathbf{T}$ in the previous equation is defined as:

$$\begin{aligned}\mathbf{T}^{(2\hat{k})}(s_k^{\pm},\cdots,s_{k+2\hat{k}-1}^{\pm}) &= \prod_{n=k}^{k+\hat{k}-1}\prod_{m=0}^{\hat{k}}\exp\left[-\hbar^{-1}(s_{n+m}^+ - s_{n+m}^-)(h_{n+\hat{k},n}s_n^+ - h_{n+\hat{k},n}^*s_n^+)\right] \\ &\times \langle s_{n+1}^+|\exp(-i\hbar^{-1}\hat{H}_s t)|s_n^+\rangle\langle s_n^-|\exp(-i\hbar^{-1}\hat{H}_s t)|s_{n+1}^-\rangle\end{aligned} \qquad (2.15)$$

The reduced density matrix $r_s$ at time $k\Delta t$ can then be recovered from $\mathbf{A}$ as:

$$r_s\left(\mathbf{s}_k^\pm; k\Delta t\right) = \mathbf{A}^{(\bar{k})}\left(\mathbf{s}_N^\pm, 0, \cdots, 0; k\Delta t\right) \exp\left[-\hbar^{-1}\left(\mathbf{s}_k^+ - \mathbf{s}_k^-\right)\left(\mathbf{h}_{nnn}\mathbf{s}_n^+ - \mathbf{h}_{nn}^*\mathbf{s}_n^+\right)\right] \qquad (2.16)$$

### III. Results

A sound understanding of the spectral properties of LHC–II cannot be accomplished without detailed knowledge of its energy eigenstates. Diagonalization of the PPC Hamiltonian revealed that while the ordering of the eigenvalues was rather sensitive to the exact energies assigned to each chlorophyll site, the character of the eigenstates were not, as seen by the coefficients of the eigenvectors in the site representation $\left|e_j\right\rangle = \sum_{k=1}^{n} c_k \left|u_k\right\rangle\left\langle u_k\right|$. Figure 2 depicts the eigenstates of the single–exciton Hamiltonian, labeled with the capital Greek letters A—Λ in increasing order of energy. All figures of LHC–II were produced with VMD[62] and rendered with POV–Ray.

Some eigenstates represent primarily localized sites, such as B (B2), Θ (A7) and I (A2). Other eigenstates represent mixing of nearest neighbors, much like the mixing of local atomic orbitals to form delocalized molecular orbitals in molecular orbital theory. The states E and M represent mixing of the A3–B3 pair, while the states H and K represent mixing of the A4–A5 pair.

Several eigenstates exhibit more extensive mixing of site character, which is most apparent in states such as Z and Λ, which consist almost entirely of the Chl *a* sites A1, B1 and A2 and are related by a sign change of the population amplitudes of the sites B1 and A2. The triplet A–Γ–Δ, representing mainly the Chl *b* sites B5, A6 and B6, also exhibits similar behavior. These states have an unusual spatial locus as they span sites both in the upper (B5) and lower (A6, B6) rings of LHC–II. While in Δ all three states have population amplitudes of the same sign, the sign of the amplitude of B6 is flipped in A, while that of A6 is changed in Γ.

The code used to generate the following results was compiled with a MIPSpro f77 compiler version 7.30 and tested on an SGI Octane 2x 300 MHz R12K server with 1280 MB of RAM. Monte Carlo calculation of the coherence length was carried out for LHC–II over a variety of simulation parameters, namely: $m$, the number of Monte Carlo iterations per dimension of integration; $N$, the number of time steps; $T$, the absolute temperature of the system in Kelvin; and $x$, the bath strength. First of all, a suitable $m$ was chosen so as to have a suitably low error statistical sampling of the coherence length, $z$, which to first order can be found to be:

$$z = 2\sqrt{\frac{N}{m}} \tag{3.1}$$

The rate of convergence was determined for the Monte Carlo simulation process for the coherence length. The coherence length was measured after an equilibration phase of equal length. A suitable value was found to be $m = 5 \times 10^5$, which was used for all subsequent simulations. For comparison, it was found that convergence to a stable and repeatable value of the coherence length required at least $m = 2 \times 10^5$ for the B800 ring from the complex LH–II of *Rs. molischianum*. ($N$=32, $x$=0, $T$=300 K).

From the Hamiltonian for the LHC–II system, it was found that the coherence length, as in the case of LH–II,[39] decreased greatly with increasing temperature. In the absence of dissipation, coherent paths seem to dominate at liquid helium temperatures (4 K) while at physiological temperatures (310 K) the coherence length was reported to be much lower. With increasing bath strength ($x$=0.0 to $x$=0.1 to $x$=1.0) it was also determined that the coherence length is dramatically decreased at all temperatures, like in LH–II, but more dramatically so. This decrease compared to LH–II may be attributed to the lower symmetry of the LHC–II complex as compared to the $C_8$ symmetry of LH–II in *Rs. molischianum*.

Due to the lower symmetry of the system, is it particularly instructive to inspect typical paths generated by the Monte Carlo process to look at specific sites accessed in the course of thermal equilibration in the LHC–II system. At physiological temperatures, the paths are essentially localized on single sites regardless of the strength of system–bath coupling, with the infrequent but notable exceptions of A2–B2 coupling and A3–B3 coupling, which occasionally persists in the regime $0 = x < 0.1$. In the light of the parameters used, this is perhaps not too surprising, as these two couplings are the largest of all such pairwise interactions.[34] (A2–B2, 121.6 cm$^{-1}$, A3–B3, 111.9 cm$^{-1}$) The experimental detection of convincing evidence for excitonic coupling due to the A2/B2 pair vindicates the results presented here, suggesting that the A2/B2 pair represents a terminal state from which inter–monomer or even inter–complex excitation energy transfer can occur.[63] However, the experimental use of two–photon excitation[64] cannot be adequately described within the scope of the one–exciton Hamiltonian due to the possibility of multi–exciton effects at the high intensities used nonlinear optical experiments.[65–67] However, the one–exciton model is still sufficient to describe light absorption under physiological conditions due to the comparatively lower intensity of ambient sunlight.

At cryogenic temperatures and in the absence of dissipative effects, paths are essentially delocalized over a substantial portion of the LHC–II monomer, while not quite completely exploring the entire domain of the complex. The introduction of the dissipative bath with system–bath couplings on the order of $x > 0.1$ leads to a dramatic decrease in the coherence length at all temperatures. Of particular interest is the dominance of a particular motif in the unusual peak in the temperature dependence plot below 50 K. In this regime, paths containing the sites B1, B5 and B6, with B6 the dominant site for most paths, were surprisingly stable to dissipative effects, persisting even for strong system–bath coupling ($x = 1$) at sufficiently low temperatures. The dominance of this path could be an artifact of the simulation process due to the lack of equilibration, even when allowed a total of $2m = 10^6$ passes per dimension. However, the persistence of such a structure could indicate the presence of a short–range, delocalized excitonic interaction at cryogenic temperatures, which is certainly encouraged by the relatively strong couplings in the model used.[34] (B1–B5, 51.6 cm$^{-1}$; B5–B6, -25.2 cm$^{-1}$) The greatly diminished dominance of this interaction at temperatures above ca. 50 K might be a factor in resolving apparent discrepancies in experiments conducted in liquid helium (4 K) versus those conducted at higher temperatures.[68]

Such paths are probably responsible for unusually high values of the coherence length in the low–temperature regime, which is unfortunately dependent on the ordering of the basis set used to construct the PPC Hamiltonian: although physically proximate to each other, the B1 site was by convention not adjacent to the B5 and B6 matrix entries. This is an unforeseen weakness of the definition of the coherence length used in this study due to the asymmetry of the system. However, other measures of coherence exist that do not exhibit this disadvantage.[69]

The site energies of the PPC Hamiltonian were varied to explore the effect of asymmetry in the site energy distribution on the coherence length. As expected, the unphysical case of having equal energies for all sites had a larger coherence length at each value of the absolute temperature sampled than for parameters corresponding more realistically to the presence of different types of chlorophylls, namely Chl *a* and Chl *b*. This, in turn, led to a larger coherence length than for sites with completely different energies. However, it can be seen that the effect of static disorder, as modeled through the choice of site energies, has little additional effect at physiological temperatures, where the vast majority of paths correspond to completely localized excitons, even in the absence of a dissipative bath. Thus, our numerical results from Monte Carlo path integration point conclusively towards the presence of excitons localized on single sites, with the occasional occurrence of excitons delocalized over the strongly–coupled site pairs A2/B2 and A3/B3. Below ca. 50 K, a spatially large delocalized path over the B1,

B5 and B6 sites was found to be a recurrent motif in the paths sampled. These three coherent subsystems are depicted in Figure 3, with clearly shows that these coherent subsystems are all located in the periphery of the LHC–II complex. This may have significant implications for inter–complex transfer of excitation energy.[24]

The next stage of our numerical investigations centered on the time evolution of the coherent subsystems identified above. For the A2/B2 and A3/B3 pairs, the Hamiltonian was reduced to that of a two–level system, e.g. with level splitting of $\Delta = 651$ cm$^{-1}$ and coupling of $c = 121.6$ cm$^{-1}$ for the A2/B2 pair, which exhibits the strongest coupling in our full system Hamiltonian. To model the Chl $b \rightarrow$ Chl $a$ excitation energy transfer process, the system was started with the initial density matrix $r'(t=0) = |u_b\rangle\langle u_b|$, where $|u_b\rangle$ denotes the Chl $b$ state. When the system is completely equilibrated, the system is expected to show Boltzmann statistics, with the probability of occupying the Chl $a$ site (which is of lower energy) at 300 K is:

$$r'_{aa}(t=+\infty) = \frac{1}{1+e^{-b\Delta}} = 0.958 \qquad (3.2)$$

To ensure the convergence of the QUAPI algorithm, the memory time $\hat{k}\Delta t$ of the A2/B2 system under the parameters T = 300 K, $x = 0.3$ and $w_c = 120$ cm$^{-1}$ was determined. The quantity $r'_{aa}$, corresponding to the population of the A2 state, was measured as a function of time to compare the effect of the step size on the time evolution of the B2 population. It was found that the memory length of the system was around 75 fs, as simulations with a fixed time step of $\Delta t = 15$ fs and increasing $\hat{k}$ were unstable in the range $1 \leq \hat{k} \leq 4$, whereas those with $\hat{k} \geq 5$ were nearly coincident, as summarized in Figure 4. Next, the minimum $\hat{k}$ necessary to fully account for non–Markovian effects was determined by fixing the memory time $\hat{k}\Delta t = 72$ fs and varying $\hat{k}$. It was found that a suitable value was $\hat{k} = 8$, as shown by the near convergence of results for $8 \leq \hat{k} \leq 9$ and the wide disparity for $1 \leq \hat{k} \leq 7$ in Figure 5.

*To be reported:*

- *Comparison of TLS numerical results with theoretical (analytic) expression?*
- *Try k_circ (k_max) = 7 (Figure 4)*
- *Determine more accurate value of memory time? 75 +/- 15 fs ok?*
- *A2/B2, A3/B3, B1-B5-B6 at 300 K, 77 K , 4 K*
- *Kinetics of thermalization of pure states (2LS and 3LS)*
- *Check sensitivity of coherent subsystems to omega_c parameter!*

*To be concluded:*

- *Resolution of QUAPI insufficient to accurately resolve coherences at 77 K due to long memory time (~75 fs) and need for high k_circ (> 6). Complexity scaling of $O(n^{\wedge}(2*k\_circ))$ prohibits the use of high k_circ.*
- *REDO abstract!*

*To check:*

- *Is the convention for forward (-) and backward (+) paths correct???*
- *Graphics come out ok? If not, will have to render at even higher resolution (currently 2000 x 2000)*
- *Formatting errors in references*

*?????*

- *I don't know why the QUAPI curves do not smoothly tend toward the origin!!!!!!!!!!!!*
- *Explain why MCPI results conflict with QUAPI results???*

## IV. Conclusion

Energy absorption by LHC–II to produce the first excited singlet state $S_1$ involves the access of $Q_y$ eigenstates. The eigenstates have been found to be mostly localized in nature, although some eigenstates exhibit some degree of coherence over spatially separated chlorophyll sites. As in earlier work employing this model,[39] thermalization is the main reason for localization of excitons at physiological temperatures. Static and dynamic disorders induce further localization but are not essential to explain decoherence of spectral eigenstates. Delocalized excitonic behavior was found only at cryogenic temperatures and with extremely weak or non–existent system–bath couplings. As similar conclusions were reached based on previous work on the B800 ring of the bacterial system LH–II of *Rs. molischianum*, this only serves to reinforce the robust and general nature of the above conclusions, despite the simplicity of the model employed. The primary difference between LHC–II and LH–II is the presence of small coherent subsystems that persist at relatively high temperatures, although at physiological temperatures no coherence was detected using the QUAPI real–time propagator algorithm. The presence of these unique features (Figure 3) can be attributed to symmetry breaking, as the symmetry of LHC–II is considerably lower than that of LH–II, resulting in the formation of strongly coupled subsystems that exhibit little or no external couplings to other system degrees of freedom.

Propagation in real time of the reduced density matrix showed the expected thermalization towards the Boltzmann–averaged density matrix. The calculated thermalization rate of ca. 4 ps establishes an upper limit on the speed of inter–monomeric excitation energy transfer as such energy transfer, which is known to be highly efficient, must occur before the exciton is dissipated into the protein environment, rendering the absorbed light energy useless.

Future dynamical work could include the study of non–photochemical quenching effects by other molecules such as carotenes and carotenoids. These represent crucial biochemical pathways by which excess excitation energy is dissipated, preventing possible photochemical damage via the formation of hazardous electronically excited side products such as singlet oxygen.

Acknowledgements: The authors thank Dr. W. Kühlbrandt for providing the draft structure of LHC–II in a digital format.

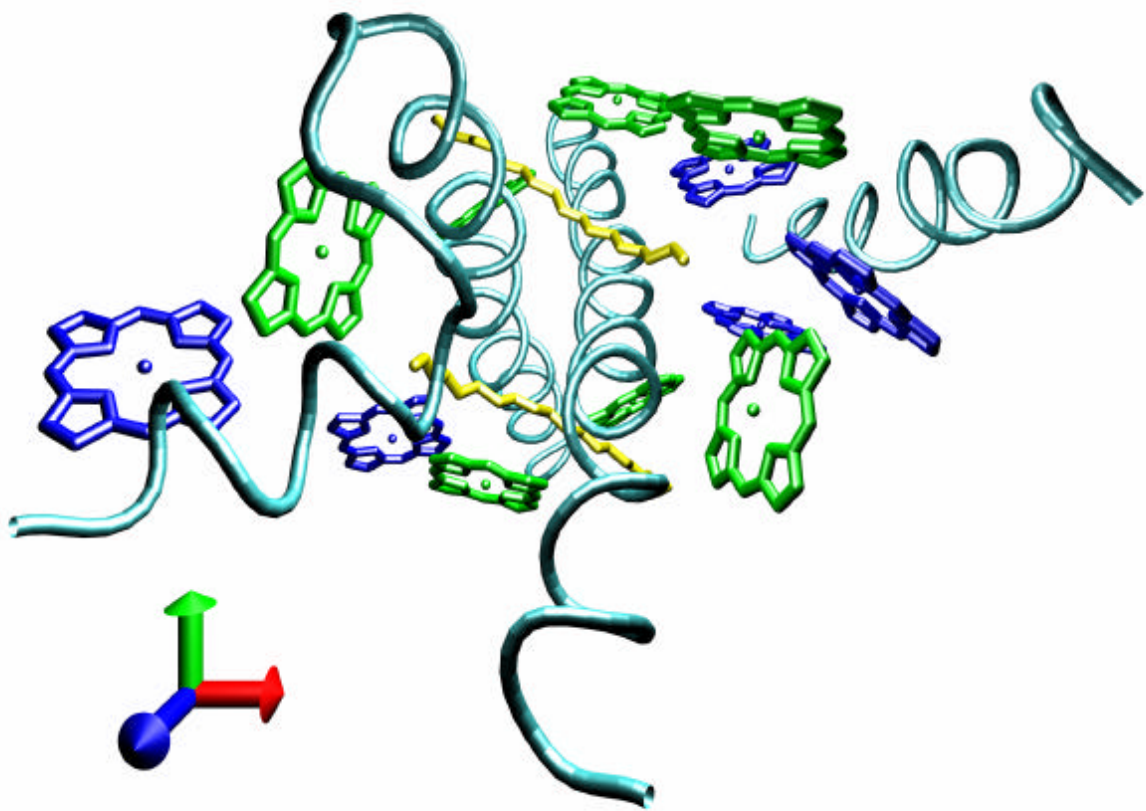

**Figure 1.** The molecular model of LHC–II employed in this work, with coordinate file courtesy of Dr. W. Kühlbrandt.[12] The protein backbone is shown in cyan, with the carbon polyene chain of two lutein moieties shown in yellow. Green tetrapyrrole rings indicate the position and orientations of chlorophyll *a* (Chl *a*) moieties, while blue rings denote chlorophyll *b* (Chl *b*) moieties. All figures of LHC–II in this work were produced with VMD[62] and rendered with POV–Ray.

**Figure 2.** Position representation of energy eigenstates, arranged from left to right, top to bottom in ascending order of energy and labeled with the capital Greek letters Α—Λ. Diagonalization was performed with an identity–differentiated Hamiltonian.[34] The protein backbone is colored brown, with the lutein chains in yellow. Chlorophyll sites with positive amplitude are colored red, while those with negative amplitude are colored blue. Sites with negligible contribution to the eigenstates are colored black.

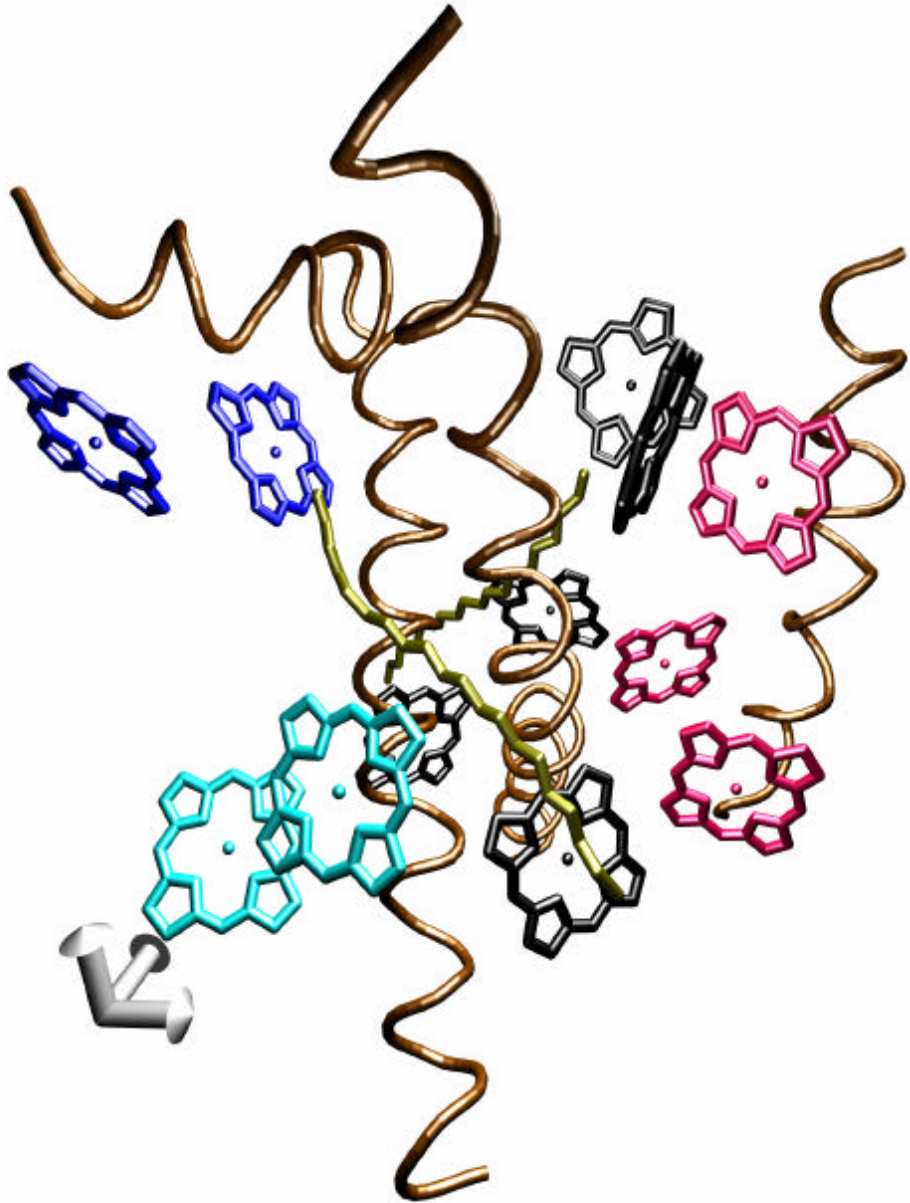

**Figure 3**. A side view of LHC–II, showing the persistent delocalized motifs in LHC–II, namely A2/B2 (cyan), A3/B3 (blue), and B1–B5–B6 (pink). The former two occasionally persist even in paths at physiological temperatures, while the last is observed only in paths sampled at a temperature below 50 K.

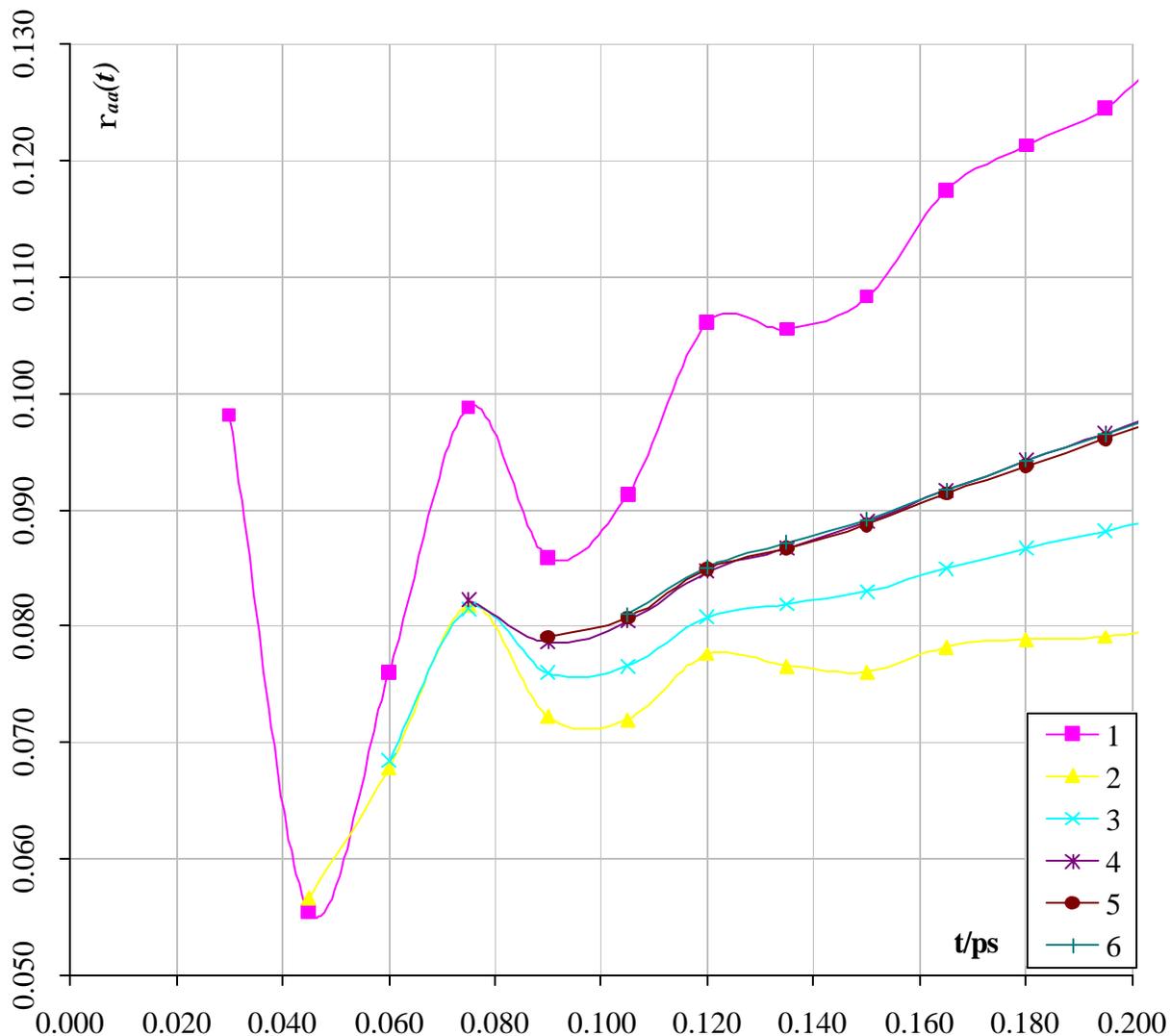

**Figure 4**. Determination of the memory length of the A2/B2 system. The diagonal element $r_{aa}$, corresponding to the population of the A2 state, was measured as a function of time to compare the effect of the step size on the time evolution of the A2 population. Symbols denote raw output data, connected by smoothed spline lines. The time step was fixed at $\Delta t = 15$ fs while the maximum temporal displacement parameter $\hat{k}$ was varied to find the memory time $\Delta t \cdot \hat{k}$. It was found that the memory time of the system was approximately 75 fs.

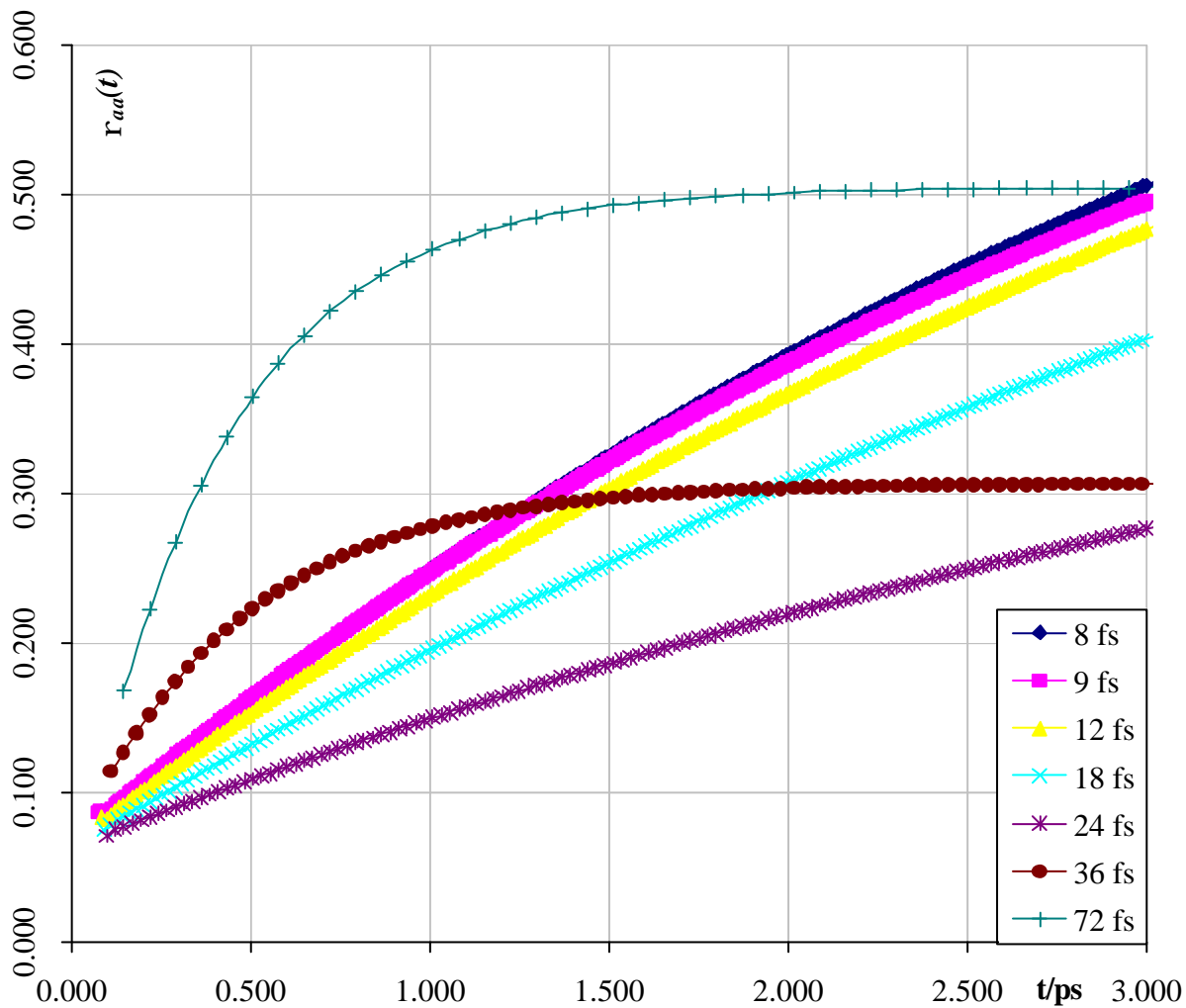

**Figure 5.** Variation of the time step $\Delta t$ and the maximum temporal displacement parameter $\hat{k}$, maintaining a constant memory time of $\Delta t \cdot \hat{k} = 72$ fs. Symbols denote raw output data, connected by smoothed spline lines. Comparing these plots show that convergence is attained with $\hat{k} = 8$.